\documentclass[useAMS,usenatbib,aas_macros]{mn2e}

\usepackage{lscape}
\usepackage{graphicx}
\usepackage{colortbl}
\usepackage{amssymb}
\usepackage{natbib}
\usepackage{times}
\usepackage{aas_macros}
\usepackage{threeparttable}

\newcommand{\msun}{~\mathrm{M}_{\odot}}
\usepackage{color}

\def\simpropto{\lower.2ex\hbox{$\; \buildrel \propto \over \sim \;$}}
\def\ltsim{\lower.5ex\hbox{$\; \buildrel < \over \sim \;$}}
\def\gtsim{\lower.5ex\hbox{$\; \buildrel > \over \sim \;$}}

\begin{document}
\title[Revised rate coefficients for H$_2$ and H$^-$]{Revised rate coefficients for H$_2$ and H$^-$ destruction by realistic stellar spectra}
\author[B. Agarwal, et al.]{Bhaskar Agarwal$^1$\thanks{E-mail:
agarwalb@mpe.mpg.de}, Sadegh Khochfar$^{2,1}$,\\
$^1$Max-Planck-Institut f{\"u}r extraterrestrische Physik,
Giessenbachstra\ss{}e, 85748 Garching, Germany\\
$^2$Institute for Astronomy, University of Edinburgh, Royal Observatory, Edinburgh, EH9 3HJ}


\date{00 Jun 2014}
\pagerange{\pageref{firstpage}--\pageref{lastpage}} \pubyear{0000}
\maketitle

\label{firstpage}


\begin{abstract}
Understanding the processes that can destroy H$_2$ and H$^-$ species is quintessential in governing the formation of the first stars, black holes and galaxies. In this study we compute the reaction rate coefficients for H$_2$ photo--dissociation by Lyman--Werner photons ($11.2 - 13.6$ eV), and H$^-$ photo--detachment by 0.76 eV photons emanating from self-consistent stellar populations that we model using publicly available stellar synthesis codes. 
So far studies  that include chemical networks for the formation of molecular hydrogen take these processes into account by assuming that the source spectra can be approximated by a power-law dependency or a black-body spectrum at 10$^4$ or $10^5$ K. We show that using spectra generated from realistic stellar population models can alter the reaction rates for photo-dissociation, $\rm k_{\rm{di}}$, and photo-detachment, $\rm k_{\rm{de}}$, significantly.
In particular, $\rm k_{\rm{de}}$ can be up to $\sim 2-4$ orders of magnitude lower in the case of realistic stellar spectra suggesting that previous calculations have over-estimated the impact that radiation has on lowering H$_2$ abundances. In contrast to burst modes of star formation, we find that models with continuous star formation predict increasing  $\rm k_{\rm{de}}$ and $\rm k_{\rm{di}}$, which makes it necessary to include the star formation history of sources to derive self-consistent reaction rates, and that it is not enough to just calculate J$_{21}$ for the background. 
For models with constant star formation rate the change in shape of the spectral energy distribution leads to a non-negligible late-time contribution to $\rm k_{\rm{de}}$ and $\rm k_{\rm{di}}$,  and we present self-consistently derived cosmological reaction rates based on star formation rates consistent with observations of the high redshift Universe.
\end{abstract}



\section{Introduction}

The first generation of stars (Population III, or Pop III) in the early universe are believed to form from  cooling and subsequent collapse of  primordial un--enriched gas \citep[e.g.][]{Tegmark:1997p937,Bromm:1999p2573}. At temperature below $\sim 8000$ K, primordial gas in the early Universe cools most efficiently via molecular hydrogen to temperatures of $\approx$ few 100 K at which point the Jeans mass can be exceeded and gravitational runaway collapse of gas can proceed \citep[e.g.][]{Omukai:1998p2622,Abel:2002p131,Yoshida:2003p51}. This channel of cooling is of importance for the formation of the first stars (Population III, or Pop III) in mini-haloes at $z \geq 10$, which kick-start the metal enrichment of the inter-galactic medium and inter-stellar medium \citep[e.g.][]{Maio:2011p104,Muratov:2012p2436}, and subsequently lead to Population II (Pop II) star formation. The complete suppression of of H$_2$ cooling has been argued to lead to the formation of black holes via the direct collapse of Jeans-unstable gas clouds of $ \sim 10^{5-6}$ M$_{\odot}$ in atomic cooling haloes  \citep[e.g.][]{Rees:1978p2743,Omukai:2001p128,Spaans:2006p58,Begelman:2006p3700,Shang:2010p33}. 

It is clear from above that the abundance of H$_2$ molecules will significantly influence structure formation in the early Universe and that accurate modelling of its formation and destruction is essential for theoretical studies that explore the same and significant work has been done in modelling the formation of H$_2 $ molecules in a cosmological context \citep[e.g.][]{Stecher:1967p3561,Allison:1970p3581,Dalgarno:1970p3590,deJong:1972p3487}.


At lower densities \citep[e.g.][]{Lepp:1984p3301,Lepp:2002p2445}, H$^-$ provides an extremely efficient channel to form H$_2$ via the reaction
\begin{eqnarray}
&\rm H^{} & +\ e \rightarrow \rm H^- + \gamma \label{reac.hmfromh} \\
&\rm H^-& +\ \rm H \rightarrow \rm H_2 + e \label{reac.h2fromhm}
\end{eqnarray}

\noindent Any mechanism that destroys H$_2$ and H$^-$ is critical to the formation of the first objects in the Universe \citep[e.g.][]{Ciardi:2000p82,Haiman:2000p87,Glover:2001p2561,Glover:2006p3695}.
Thus, photo--dissociation of H$_2$ by Lyman--Werner (LW) photons ($h\nu = 11.2 - 13.6$~eV), and photo--detachment of H$^-$ by $>$ 0.76 eV photons must be accounted for.
\begin{eqnarray}
&\rm H_2& +\ \gamma_{LW} \rightarrow \rm H + H \label{reac.pdi} \\
&\rm H^-& +\ \gamma_{0.76} \rightarrow \rm H + e \label{reac.pde}
\end{eqnarray}

The reaction rate coefficients for the destruction of these two species have been derived in the literature by assuming that sources have spectra that can be approximated as power--law or a black--body type \citep[e.g.][]{Draine:1996p2556,Abel:1997p3456,Galli:1998p3284,Galli:1998p3288,Omukai:2001p128}. However, 
the spectral energy distribution (SED) of stars shows a strong time dependence in its normalisation and shape at the relevant energies  \citep[e.g.][]{Leitherer:1999p112,Schaerer:2002p21}. While the former is generally accounted for \citep[e.g.][]{Dijkstra:2008p45,Agarwal:2012p2110,Visbal:2014p3754,Latif:2014p3556}, the latter has been completely neglected so far. 
Previous estimates have been based on thermal spectra, mostly at a temperature of 10$^4$~K (T4) that is argued to mimic a Pop II-type stellar population, and at $10^5$~K (T5) that mimics a Pop III type population \citep{Omukai:2001p128,Shang:2010p33,WolcottGreen:2011p121}. 
The aim of this paper is to derive reaction rate coefficients for the destruction of H$_2$ and H$^-$, based on stellar synthesis models representative of stellar populations in galaxies and to investigate their impact on the formation and destruction of H$_2$ molecules. 

The paper is organised as follows. We will outline the methodology in Sec. \ref{sec.methodology}, followed by the results that are presented in Sec. \ref{sec.results}. Finally, the conclusions and a critical discussion of the implication of the work presented in this study is presented in Sec. \ref{sec.conclusions}.



\begin{table*}
\centering
\caption[1.8\columnwidth]{Properties of the stellar populations analysed in this work and their corresponding references. The IMF slope is of the form $\xi \propto m^{-\alpha}$, and the mass interval is listed in the form [M$_{min}$, M$_{max}$]. The total stellar mass is proportional to the stellar age in the case of a continuous SFR mode, and for the burst modes, the SFR is assumed to occur in bursts at each time interval.}
\begin{threeparttable}
\begin{tabular*}{2\columnwidth}{@{\extracolsep{\fill}}cccccccc}
\hline
&Ref. & IMF & slope & Mass Interval&M$_{\rm tot}$ & SFR & Name \\
 &  &  & &[$\msun$] & [$\msun$] &[$\msun/\rm yr$] & \\
 \\ [-1.5ex] \hline \\ [-1.5ex]
Pop III & ST99\tnote{a} & Salpeter & 2.35 & 50,500& $10^6$& burst & ST99\_III\\
Pop II&ST99\tnote{b} & Chabrier & 2.3,2.3 & [0.1,1],[1,100]& $10^6$& burst & ST99\_Cb \\
Pop II & ST99 & Salpeter & 2.35 & 1,100& $\propto t$ &1&ST99\_Sc\\
Pop II & ST99 & Salpeter & 2.35 & 1,100& $10^6$ & burst &ST99\_Sb\\
 \\ [-1.5ex] \hline \\ [-1.5ex]
\end{tabular*}
\begin{tablenotes}
\item[a] Code was modified to include the modified IMF (personal communication with Daniel Schaerer)
\item[b] Code was modified to include the IMF and a lower metallicity limit
\end{tablenotes}
\label{tab.1}
\end{threeparttable}
\end{table*}

\section{Methodology}
\label{sec.methodology}
The chemo--thermodymanic evolution of primordial gas subject to Lyman-Werner radiation \footnote{Strictly speaking Lyman-Werner radiation only corresponds to photons with energies between $11.2-13.6$ eV. However, for the sake of convenience in the following we will refer by { \it Lyman-Werner radiation} or {\it radiation } to photons responsible for photo-detachment of H$^-$ as well.}
 is governed by the reactions in Eq. \ref{reac.hmfromh} -- \ref{reac.pde}. The main coolant at low temperatures are H$_2$ molecules whose abundance is given by the corresponding reaction rates. It is common to define an effective formation rate coefficient k$_{\rm form}$ \citep{Omukai:2001p128}: 
\begin{equation}
{\rm k_{form}} \equiv {\rm k_1} \frac{{\rm k_2} n_{\rm{H}}}{n_{\rm{H}} {\rm k_2}+ {\rm k_{\rm{de}}}}
\label{eq.netform}
\end{equation}
The hydrogen number density is $n_H$ and the indices of the individual rate coefficients k$_i$ refer to the equation number of the reaction shown above. Note that it is common practice to use the modified reaction rate coefficient for H$^-$ photo-detachment ${\rm {k}_{\rm{de}}} \equiv f_{\gamma_{0.76}} n_{\rm{H}} \rm k_4 $ \citep[e.g.][]{Omukai:2001p128}.
Combining the formation and destruction processes in Eqs. \ref{reac.hmfromh} -- \ref{reac.pde} yields the following equilibrium solution for the abundance of H$_2$ molecules:
\begin{equation} \label{h2frac}
 f_{\rm{H}_2} \equiv \frac{n_{\rm{H}_2}}{ n_{\rm{H}}}  = {\rm \frac{ k_{form}}{k_{\rm{di}}}} f_e  n_{\rm{H}}
\end{equation}
with $f_e \equiv  n_{e} / n_{\rm{H}}$ as the electron number density in units of the Hydrogen number density and the modified H$_2$ photo-dissociation reaction rate coefficient k$_{\rm{di}} \equiv f_{\gamma_{LW}} n_{\rm{H}} \rm k_3 $. 
The general assumption in the literature is that k$_{\rm{de}}$ and k$_{\rm{di}}$, and thus the ratio $\rm k_{\rm{form}} / k_{\rm{di}}$ is constant with time, which is justified for fixed black body and power-law spectra. However, for SEDs from stellar populations these quantities are a strong function of stellar age as we will show later in Sec. \ref{sec.results}.   
It is useful to define the reaction coefficients k$_{\rm{de}}$ and k$_{\rm{di}}$ in terms of their SED dependence via

\begin{equation}
\rm k_{di} = \kappa_{di}\beta J_{21} \rm \ 
\label{eq.h2}
\end{equation}

\begin{equation}
\rm k_{de} = \kappa_{de}\alpha J_{21} \rm \ 
\label{eq.hm}
\end{equation}

\noindent The rate constants $\kappa$ are defined for a flat spectrum with no frequency or time dependence and carry the units of s$^{-1}$. 
The dimensionless parameter, J$_{21}$, is defined as the specific intensity at 13.6 eV, normalised to 10$^{-21}$ erg/s/cm$^2$/Hz/sr, for a point source at a distance $d$ (in cm).
 i.e.:

\begin{eqnarray}
{\rm{J}}_{21} \equiv \frac{L_{13.6}}{4\pi\ 4\pi d^2}\ {\rm erg/s/cm^2/Hz/sr} \\ \nonumber
\times \frac{1}{10^{-21} {\rm\ erg/s/cm^2/Hz/sr}}. \nonumber
\label{eq.J}
\end{eqnarray}

It becomes useful to define a normalised spectrum, L$_n$, 

\begin{equation}
{\rm L}_n = \frac{{\rm L}_\nu}{\rm L_{13.6}} \times 10^{-21}\ {\rm erg/s/cm^2/Hz/sr}
\label{eq.normsed}
\end{equation}

which is the ratio of the spectrum L$_\nu$ to its value at 13.6 eV, in the units of $10^{-21}\ {\rm erg/s/cm^2/Hz/sr}$.

The \textit{spectral parameter}, $\beta$ ($\alpha$), introduces the sensitivity of the reaction rate on the SED of the source with respect to a source with a flat spectrum.

With above definitions the time dependence of the reaction rates on the SED is split up in two components: one that is equal for both (J$_{21}$) and just reflects the amount of UV photons depending on the age of the stellar population, and the other ($\alpha$ and $\beta$, respectively) that is sensitive to the change in the shape of the SED. The constants $ \kappa_{\rm de}$ and $\kappa_{\rm di}$ carry the units of s$^{-1}$, and reflect the efficiency of the reactions.

\subsection{H$_2$}
Molecular hydrogen, in its ground state, is photo--dissociated by the two--step Solomon process \citep{Stecher:1967p3561} where absorption of LW photons, leads to electronically and vibrationally excited states that subsequently decay into the ground state. The rate constant can be derived by summing over all the bands, the product of the strength of the oscillations in a given band \citep{Allison:1970p3581} and the probability of decay \citep{Dalgarno:1970p3590}. We refer the reader to \cite{Abel:1997p3456} for a derivation of the same, where the authors derive the photodissociation rate constant $\kappa_{27} \sim 1.1 \times 10^{8}$ s$^{-1}$, for a given value of an un--normalised flux ($\rm F$ in units of erg/s/Hz/cm$^2$). Thus, for our definition of $\rm J_{21}$, $\kappa_{\rm di} =  4\pi10^{-21}\kappa_{27} = 1.38\times 10^{-12}$ s$^{-1}$ \citep[see][]{Yoshida:2003p51}. Thus Eq. \ref{eq.h2} takes the form 

\begin{equation}
\rm k_{di} = 1.38 \times 10^{-12}\beta J_{21} \ 
\label{eq.finalbeta}
\end{equation}

Here, $\beta$ is a dimensionless parameter introduced to capture the dependance of the shape of the input spectrum and is defined as the ratio of the normalised intensity of a given spectrum in the LW band to that at $\rm L_{\rm n} =1 \times 10^{-21}$ erg/s/cm$^2$/sr, i.e.

\begin{equation} 
\beta =\rm \frac{\int\limits^{\nu_{13.6}}_{\nu_{11.2}}L_{n}d\nu}{\Delta \nu_{\rm LW}} 
\label{eq.beta}
\end{equation}

where ${\nu_{11.2}}\ \&\ {\nu_{13.6}}$ denote the frequency limits corresponding to 11.2 and 13.6 eV respectively (i.e. the LW band), and $\Delta \nu_{LW} = {\nu_{13.6}} - {\nu_{11.2}}$. Using the above definition gives $\beta \approx 3, 0.9$ for T4 and T5 thermal spectra, as quoted in \cite{Shang:2010p33, Omukai:2001p128}. 

\subsection{H$^-$}
The abundance of H$^-$ plays a critical role in determining the net formation rate of molecular hydrogen, as shown in Eq.~\ref{eq.netform}.
The reaction rate for H$^-$ photo--detachment can be defined as in \cite{Omukai:2001p128}, and can be thought of as an effective cross section weighted by the irradiating spectrum defined in terms of L$_{n}$.

\begin{equation}
{\rm k_{\rm de}} = \int\limits_{\nu_{0.76}}^{\nu_{13.6}} \frac{4\pi L_n}{h\nu}\sigma_{\nu} d\nu \ (\rm s^{-1})
\label{eq.kappa_de}
\end{equation}

where ${\nu_{0.76}}\ \&\ {\nu_{13.6}}$ denote the frequency limits corresponding to 0.76 and 13.6 eV respectively and $\sigma_{\nu}$ represents the cross section, defined as

\begin{equation}
\sigma_{\nu} = 10^{-18}\lambda^3(\frac{1}{\lambda} - \frac{1}{\lambda_o})^{1.5}\sum\limits_{n=1}^{6}\rm C_n[\frac{1}{\lambda} - \frac{1}{\lambda_o}]^{\frac{n-1}{2}}\ cm^2
\end{equation}
where $\lambda_o\approx1.6 \mu m$ (0.76eV) and the wavelength terms are expressed in $\mu m$, and the parameter $\rm C_n$ is tabulated in \cite{John:1988p3398}.\footnote{We also refer the reader to the formulation presented in \cite{deJong:1972p3487,Wishart:1979p3403,Tegmark:1997p937,Abel:1997p3456}. All the above formulations agree to within percentage level of each other.}

Defining $\kappa_{de}$ as the value of k$_{\rm de}$ at J$_{21}=1$, i.e. ${\rm L}_n = 1 \times 10^{-21}$ erg/s/cm$^2$/sr, constant for all frequencies and in time  (Omukai 2001, Appendix A), the integral in eq.~\ref{eq.kappa_de} can be solved to give $\kappa_{\rm de} = 1.1\times 10^{-10}$ s$^{-1}$. 

Thus for any given normalised spectrum L$_n$, Eq.~\ref{eq.hm} can be written as

\begin{equation}
\rm k_{de}=1.1\times10^{-10}\alpha J_{21}\ 
\label{eq.finalalpha}
\end{equation}

where $\alpha$ is introduced to capture the sensitivity of the reaction rate to the shape of the input spectrum, i.e. $\alpha = 1$ for ${\rm L}_n = 1 \times 10^{-21}$ erg/s/cm$^2$ /sr. Or in other words

\begin{equation}
\alpha =\frac{\int\limits_{\nu_{0.76}}^{\nu_{13.6}} \frac{4\pi {\rm L}_{n}}{h\nu}\sigma_{\nu} d\nu}{\int\limits_{\nu_{0.76}}^{\nu_{13.6}} \frac{4\pi10^{-21}}{h\nu}\sigma_{\nu} d\nu} = \frac{\int\limits_{\nu_{0.76}}^{\nu_{13.6}} \frac{4\pi {\rm L}_{n}}{h\nu}\sigma_{\nu} d\nu}{1.1\times10^{-10}} 
\label{eq.alpha}
\end{equation}

Using the above formulation, one obtains $\alpha \approx 8$ (2000) for a power-law type (T4) spectrum as shown by Omukai (2001).

\begin{figure*}
\centering
\includegraphics[width=1.5\columnwidth,]{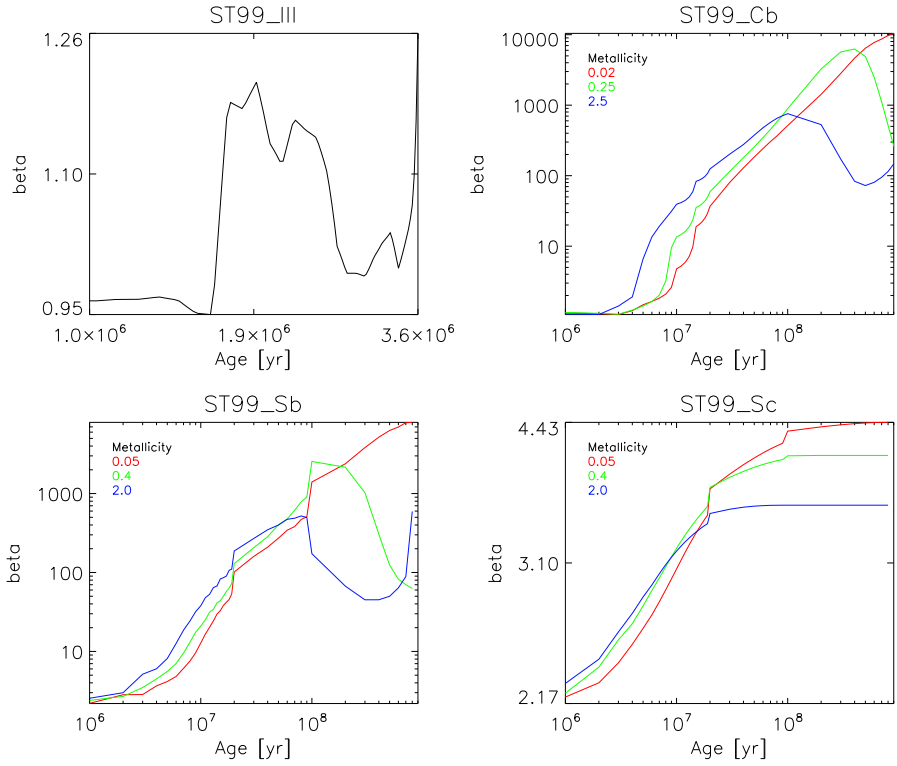}
\caption[beta for H$_2$]{The $\beta$ parameter for H$_2$ dissociation from the different stellar populations analysed in this study. Note that the metallicities are quoted in solar units with $Z_{\sun} =0.02$. In the bottom right panel, a continuous star formation rate of $1\msun \rm / yr$ is employed, whereas in the rest of the panels a single burst of $10^6\msun$ is used to model the SFR (see Table.~\ref{tab.1}). }
\label{fig.beta}
\end{figure*}

\begin{figure*}
\centering
\includegraphics[width=1.5\columnwidth,]{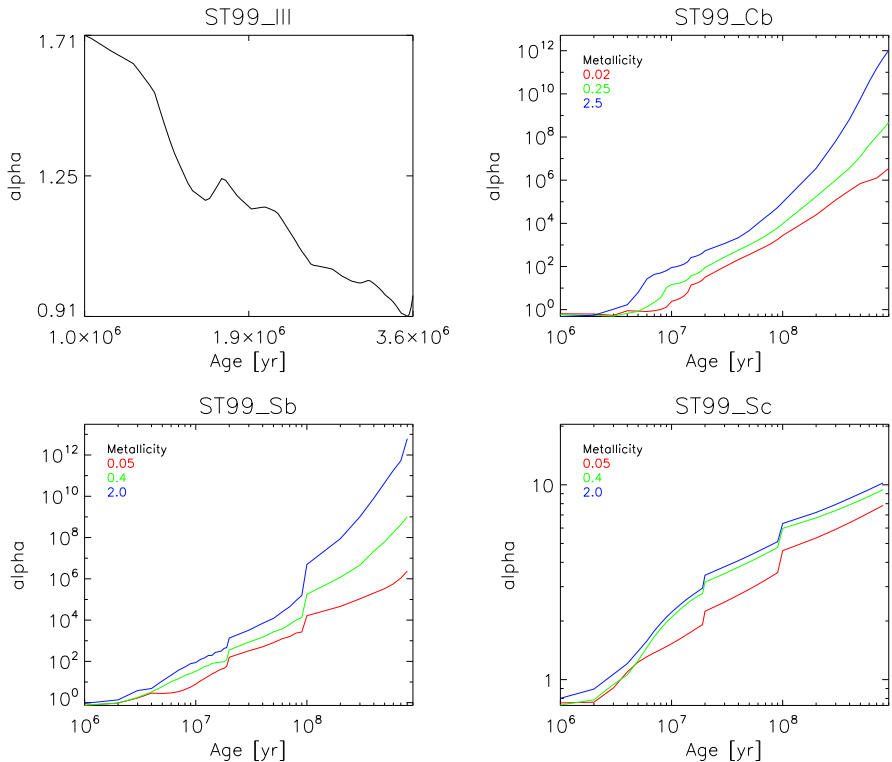}
\caption[alpha for H$^-$]{The $\alpha$ parameter for H$^-$ photo-detachment from different stellar populations analysed in this study (see caption for Fig.~\ref{fig.beta} for more details).}
\label{fig.alpha}
\end{figure*}

\begin{figure*}
\centering
\includegraphics[width=1.5\columnwidth,]{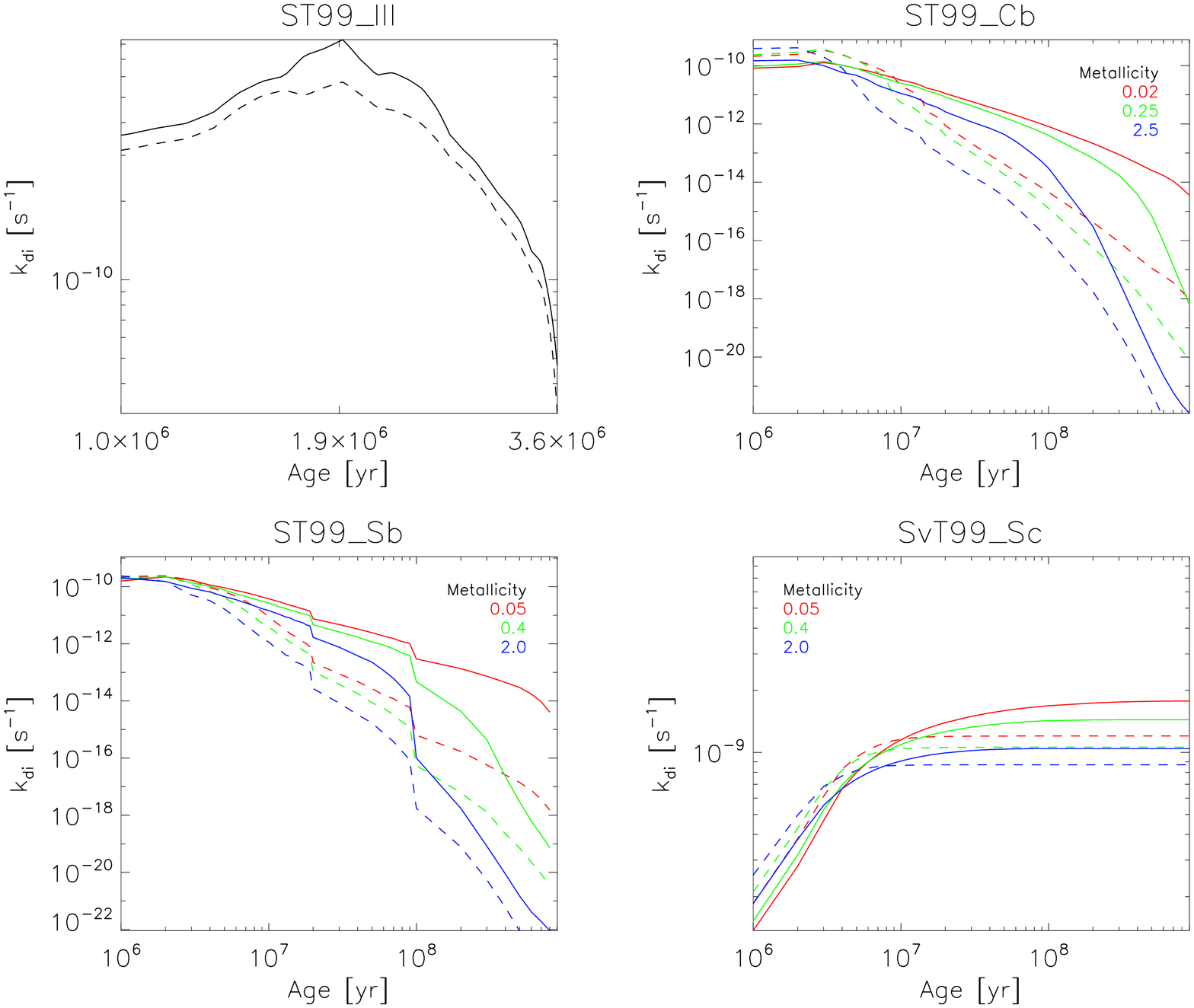}
\caption[reaction rate coefficient H$_2$]{The reaction rate coefficient k$_{\rm di}$ for H$_2$ photo--dissociation computed at a distance of 5 kpc (physical) from a given stellar population (see caption for Fig.~\ref{fig.beta} for more details). The dashed lines are computed for $\beta = 0.1$ (top left panel), and $\beta=3$ (rest) corresponding to a T5 and T4 spectrum respectively, with the same J$_{21}$ for the stellar populations.}
\label{fig.betaj21}
\end{figure*}

\begin{figure*}
\centering
\includegraphics[width=1.5\columnwidth,]{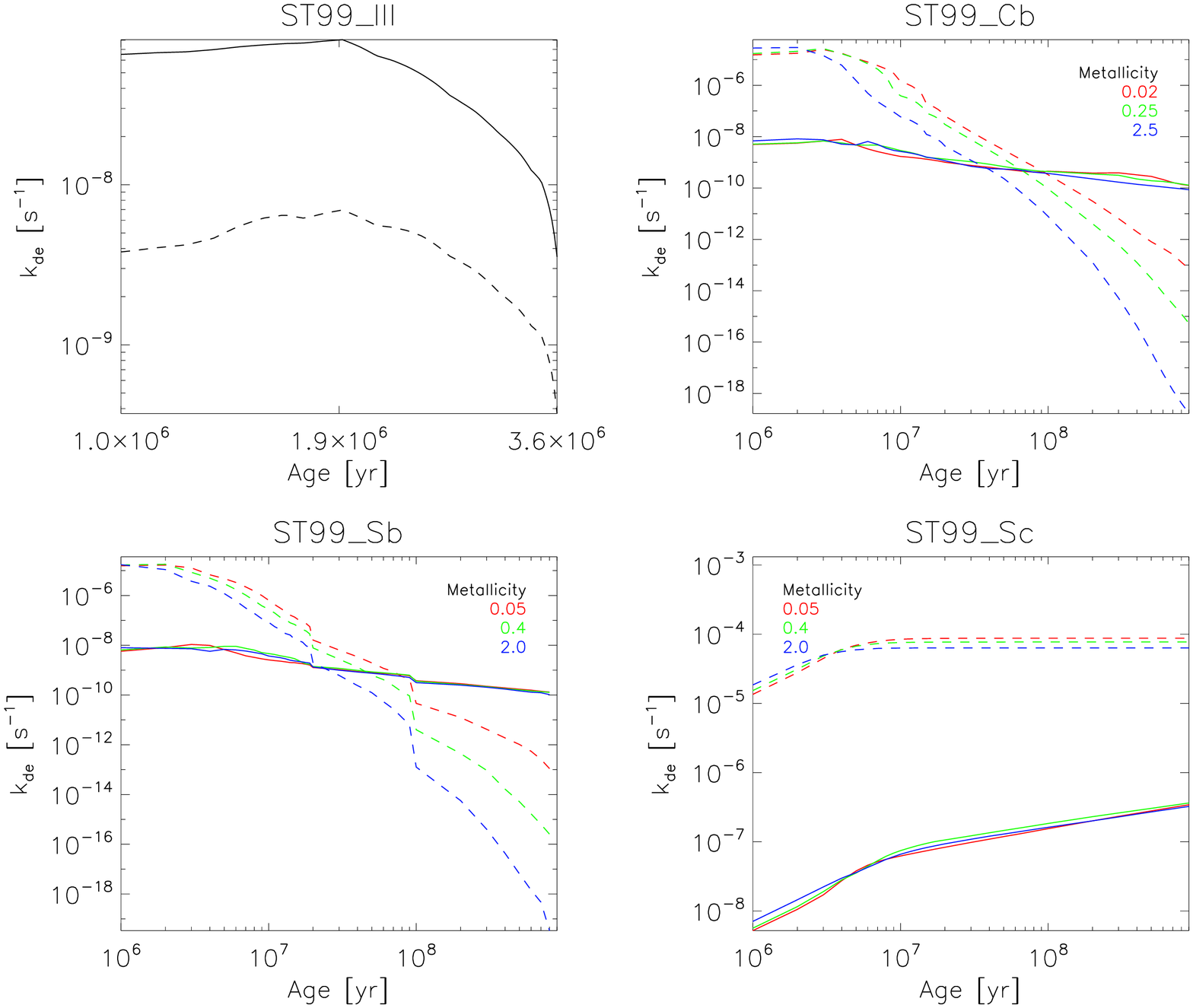}
\caption[reaction rate coefficient H$^-$]{The reaction rate coefficient k$_{\rm de}$ for H$^-$ photo--detachment computed at a distance of 5 kpc (physical) from a given stellar population (see caption for Fig.~\ref{fig.beta} for more details). The dashed lines are computed for $\alpha = 8$ (top left panel), and $\alpha=2000$ (rest) corresponding to a T5 and T4 spectrum respectively, with the same J$_{21}$ derived from the stellar populations.}
\label{fig.alphaj21}
\end{figure*}

\subsection{Stellar populations}
We use the publicly available stellar synthesis code {\sc starburst 99} \citep{Leitherer:1999p112}, and \cite{Schaerer:2002p21} to model Pop II and Pop III type stellar populations with different star formation histories. The properties of each of the models analysed in this study are listed in Table~\ref{tab.1}. 
In brief, we generate Pop II SEDs for either Salpeter or Chabrier IMFs using single burst models that form $10^6$ M$_{\odot}$ instantaneously, or continuous star formation models with 1 M$_{\odot}/$yr. In addition, we model Pop III SEDs using a Salpeter IMF with a high mass end of 500 M$_{\odot}$. \footnote{The IMF of Pop III stars is heavily debated at this point going from flat to top-heavy \citep[e.g.][]{Hirano:2014p3818}.}

\section{Results}
\label{sec.results}
Here we present the results of our implementation. We find that the spectral parameters are extremely sensitive to the age of the stellar population that is producing the photons.\footnote{Tabulated spectral indices for the curves shown in Figures. \ref{fig.beta} and \ref{fig.alpha}  are downloadable as data--tables with the source.} We will also discuss in details the impact that these spectral parameters have on the final reaction rates, and the interplay with J$_{21}$.

%
%
%
\subsection{Pop III}
For our case with the Pop III--type population (ST99\_III) where a single burst of SFR is used, we find little variation in the value of the spectral parameters over the age of the stellar population considered, as seen in Fig.~\ref{fig.beta} \& \ref{fig.alpha}. For our mass range of Pop III stars considered, $50-500\msun$, the age of the population extends over $\approx 1-3.6$~Myr only. 
Although the spectral parameters do not vary as dramatically as in the Pop II--type cases (see Figures \ref{fig.beta} \&  \ref{fig.alpha}), in the first $10^6$ yr ST99\_III produces higher values for k$_{\rm di}$ and  k$_{\rm de}$ then a comparable burst for Pop II stars as shown in Fig.~\ref{fig.betaj21} \& \ref{fig.alphaj21}. This does not come as a surprise given that the number of massive stars is larger for our choice of Pop-III IMF.

\subsection{Pop II}

Models with bursts show a stronger dependence of the spectral parameters on the age of the population beyond a few Myr (see ST99\_Cb \& ST99\_Sb in Fig.~\ref{fig.beta}), than the ones with a continuous mode of SF. Initially the corresponding bands evolve similar to the band at 13.6 eV. However, after a few million years the output at this wavelength drops more severely and we find a several order of magnitude increase in $\beta$ and $\alpha$. The time dependence of $\alpha$  is stronger than that of $\beta$ because the change in the shape of the SED with respect to 13.6 eV is stronger at 0.76 eV than in the LW-bands.  While these changes are significant one needs to calculate k$_{\rm de}$ and k$_{\rm di}$ to gauge the impact on the reaction rates, as J$_{21}$ is an equally strong function of time in the opposite direction to $\alpha$ and $\beta$, and could potentially compensate for any time evolution. 
In figures~\ref{fig.betaj21} \& \ref{fig.alphaj21} we show $\rm k_{de}$ and $\rm k_{di}$ as a function of time for a source located 5 kpc  (physical) away for different metallicities.
This involves computing J$_{21}$ for a given age and metallicity for the underlying stellar population, normalised to either $10^6\msun$ or a SFR of $1 \msun\ \rm yr^{-1}$.  Dashed lines are for reaction rates computed using the spectral parameters for a T5 spectrum for the top left panel, meant to mimic a Pop III type population, and T4 for the rest of the panels, meant to represent Pop II stars, with the same J$_{21}$ derived for the corresponding stellar models.

As expected, the impact of using SEDs from stellar population models is much stronger for the photo--detachment than for photo--dissociation. This is because the temporal evolution of the SED is generally normalised to 13.6 eV which is in the LW-band, thus making the effect less pronounced for k$_{\rm di}$. For both the burst and continuous modes, within the first $\sim$ 10 Myr, $\rm k_{di}$ is only a factor of few higher in the T4 case, after which it falls $\sim 1-2$ orders of magnitude below the realistic spectra. However, by that time the UV-output in terms of J$_{21}$ has dropped off significantly for the bursty models, suggesting that H$_2$ photo-dissociation will be mostly affected in this case within the first 10 Myr.
However, for continuously star forming galaxies the maximal destruction of H$_2$ occurs after $ \sim 10$ Myr and the black-body predictions are similar to that of the stellar SEDs. Galaxies with constant or even increasing star formation histories as predicted by galaxy formation models \citep[e.g.][]{Khochfar:2011p840,Finlator:2011p3819} will thus have H$_2$ photo-dissociation occurring more efficiently as predicted by simple black-body spectra.


The T5 spectrum is over an order of magnitude efficient in photo--detaching H$^-$ as compared to the Pop III type model analysed in this study. The T4 spectra however, produce a k$_{\rm de}$ that is more than three orders of magnitude higher than the burst modes, but only at stellar ages of $<50$ Myr, after which the T5 spectrum drops off steeply. This accentuates the role of older stellar populations if one considers bursty SF modes, as the drop in k$_{\rm de}$ between 1 Myr and 1 Gyr for the burst modes analysed here is only about two orders of magnitude, but is over 13 orders of magnitude for the T4 spectrum. However, as compared to the case with a continuous SFR, the T4 spectrum results in a detachment rate that is $\sim$ 2--3 orders of magnitude higher for the entire age of the stellar population. The implications of this deviation of the detachment rates can be severe for processes like DCBH formation, where k$_{\rm de}$ plays a pivotal role in determining the critical level of LW specific intensity needed from Pop II stars to fully suppress H$_2$ cooling.


\subsection{Inexistence of a universal J$_{\rm crit}$}

The collapse and thermodynamic fate of pristine gas is governed by the reactions discussed in this study (Eq. \ref{eq.finalbeta} and \ref{eq.finalalpha}), which are extremely sensitive to the age and SFH of the stellar population in galaxies producing the photons, and the level of J$_{21}$ as shown above. Thus, merely knowing the value of J$_{21}$, which is calculated at 13.6 eV only, is not sufficient to predict if the gas is free of molecular hydrogen.
Even for the case of suppression of Pop III star formation by a global LW background \citep[e.g.][]{Machacek:2001p150,Glover:2001p2561,OShea:2008p41}, the  time dependency on the spectral parameters $\alpha$ and $\beta$ must be accounted for.

We argue that the most drastic outcome of our work is the possibility of the absence of a universal J$_{\rm crit}$ required for direct collapse. Previous studies have shown that the value of J$_{\rm crit}$ is extremely sensitive to the rate of reactions \ref{reac.pdi} and \ref{reac.pde}  at number densities of $< 10^4$ cm$^{-3}$ \citep{Omukai:2001p128,Shang:2010p33,WolcottGreen:2011p121}, where a T4 or T5 type irradiating source was assumed. However, a given value of J$_{21}$ could be produced by stellar populations with various combinations of their properties such as age, metallicity, stellar mass, star formation rate etc.. Depending on the combination of these variables, one could obtain a large range for the value of the spectral parameters (see Fig. \ref{fig.beta} and \ref{fig.alpha}), and the corresponding value for the reaction rates \textit{for a fixed J$_{21}$}. Based on this we argue that the net formation rate of H$_2$, k$_{\rm form}$ (see Eq. \ref{eq.netform}), is a function of the stellar age, metallicity and  SF mode. Thus for a given value of k$_{\rm form}$ (or k$_{\rm de}$ and k$_{\rm di}$) there is a degeneracy in the parameter space $[(M_*, \dot{M_*}),\ Z_{\sun},\ t]$ that determines the value of the reaction rates.

\begin{figure*}
\centering
\includegraphics[width=1.75\columnwidth,]{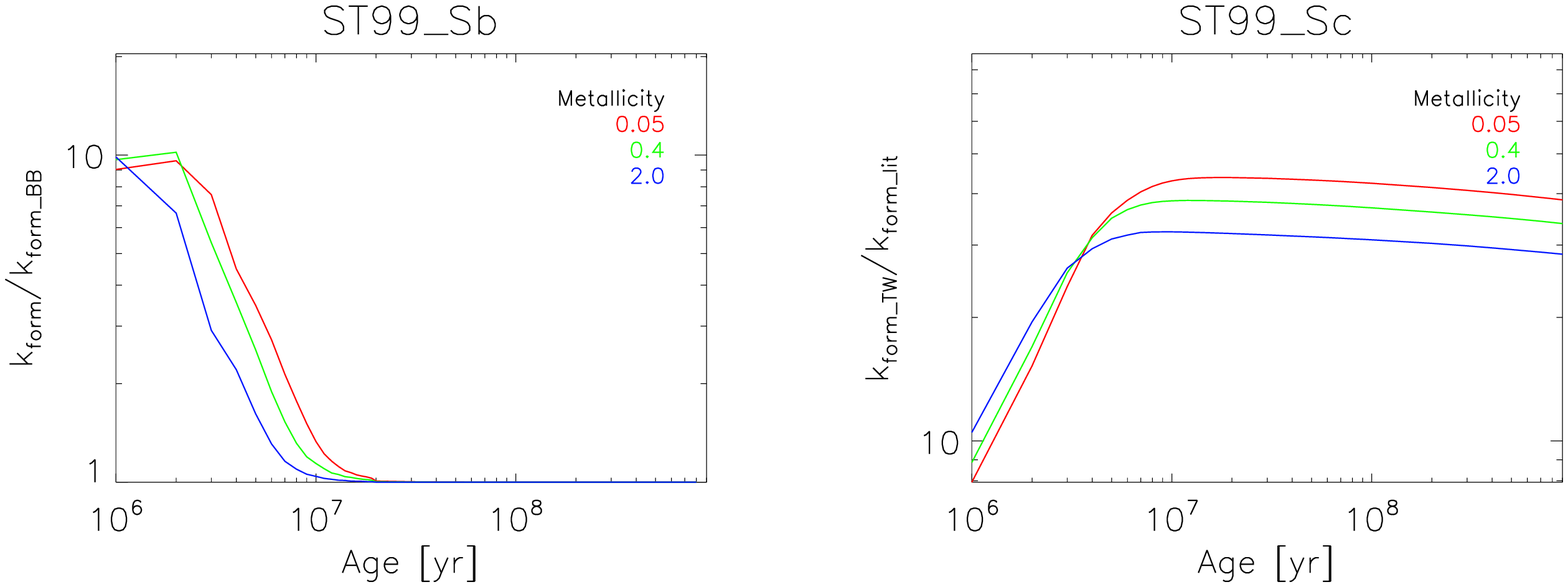}
\caption[Ratio of k$_{\rm form}$]{Ratio of k$_{\rm form}$ derived in our work to that in the literature for a T4 spectrum, assuming $\rm n_{H}= 10^3\ cm^{-3}$, and for the same value of J$_{21}$ used in Fig. \ref{fig.betaj21} and \ref{fig.alphaj21}. The ratio being larger than unity demonstrates that the spectra analysed in this work are less efficient in suppressing H$_2$ formation, especially at low densities via the H$^-$ channel.}
\label{fig.kformratio}
\end{figure*}

In Fig. \ref{fig.kformratio}, we plot the ratio of the formation rate of H$_2$ computed on the basis of this work (Eq. \ref{eq.netform}), to the values derived for a T4 spectrum, k$_{\rm form\_BB}$. The same values of J$_{21}$, computed as a function of age for the corresponding stellar model, are used for both k$_{\rm form}$ and k$_{\rm form\_BB}$, but $\alpha = 2000$ is used for the latter.

\begin{eqnarray}
&\rm k_{form}& = \rm k_{\ref{reac.h2fromhm}}\frac{k_{\ref{reac.hmfromh}}n_{H}}{k_{\ref{reac.hmfromh}}n_{H} + k_{de}}, \\ \nonumber
&\rm k_{form\_BB}& = \rm k_{\ref{reac.h2fromhm}}\frac{k_{\ref{reac.hmfromh}}n_{H}}{k_{\ref{reac.hmfromh}}n_{H} + k_{\rm de,BBT4}}
\end{eqnarray}

where we set $\rm n_{H} = 10^3\ \rm cm^{-3}$, k$_{\ref{reac.h2fromhm}}$ and k$_{\ref{reac.hmfromh}}$ are computed on the basis of \cite{Shang:2010p33}, k$_{\rm de,BBT4} = 1.1\cdot 10^{-10}\times 2000\ \rm J_{21}$, and k$_{\rm de}$ is computed on the basis of the work presented here.

The deviation of the ratio from unity arises due to the dependence of k$_{\rm de}$ on stellar age that we have demonstrated in this study, a variation that has been previously overlooked by assuming a constant value of the photo--detachment rate (k$_{\rm de,BBT4}$) for T4 type spectra. We find that for a given value of J$_{21}$, H$_2$ forms more efficiently in the case of the spectra considered in this study, hinting at the need for a higher J$_{\rm crit}$.

\subsection{Implementation}

In the derivation of the spectral parameters and reaction rates, the definition of L$_n$ and J$_{21}$ must be consistent with each other in order to avoid spurious results, i.e. difference of over 5 orders of magnitude in the reaction rates. The difference can be attributed to the fact that L$_{\nu}$ is not flat in the LW range for realistic spectra, such as the ones considered in this study. For instance, in our calculations, L$_n$ is the spectrum normalised to its own value at 13.6 eV, thus J$_{21}$ is defined \textit{at} 13.6 eV as well, and any other definition of J$_{21}$ would lead to inconsistencies. 

Also, the results presented here are either for a bursty or continuous mode of SFR. If the reader choses to implement our results, then the rates defined in Eq.~\ref{eq.finalbeta} \& \ref{eq.finalalpha} need to be multiplied by a factor of $\psi=\frac{M_*}{10^6\msun}$ or $\psi=\frac{\dot{M}_*}{1\msun/\rm yr}$, where $M_*$  and $\dot{M}_*$ denote the user specified stellar mass and SFR that would be an attribute of their underlying stellar population that mimics our model(s), i.e.

\begin{eqnarray}
&\rm k_{di}& = \kappa_{di}\beta J_{21} \rm \psi \ (s^{-1}) ,\\
\label{eq.implementkdi}
&\rm k_{de}& = \kappa_{de}\alpha J_{21} \rm \psi \ (s^{-1}) .
\label{eq.implementkde}
\end{eqnarray}

\subsection{Cosmic values for $\rm k_{di}$ and $\rm k_{de}$}

We compute the redshift evoution of the reaction rates k$_{\rm de}$ and k$_{\rm di}$, using the approach outlined in this work (i.e. dependence on metallicity, stellar age, SF mode), and on the basis of the fiducial model of \cite{Agarwal:2012p2110} which reproduces the observed cosmic star formation rate in the high redshift Universe. For their fiducial case and assuming that the Pop II stars in their work have $Z = 0.05\ Z_{\odot}$, we use their their modelled star formation histories of individual galaxies and the ST99\_Sb case to compute the reaction rates. 

\begin{figure}
\centering
\includegraphics[width=\columnwidth,]{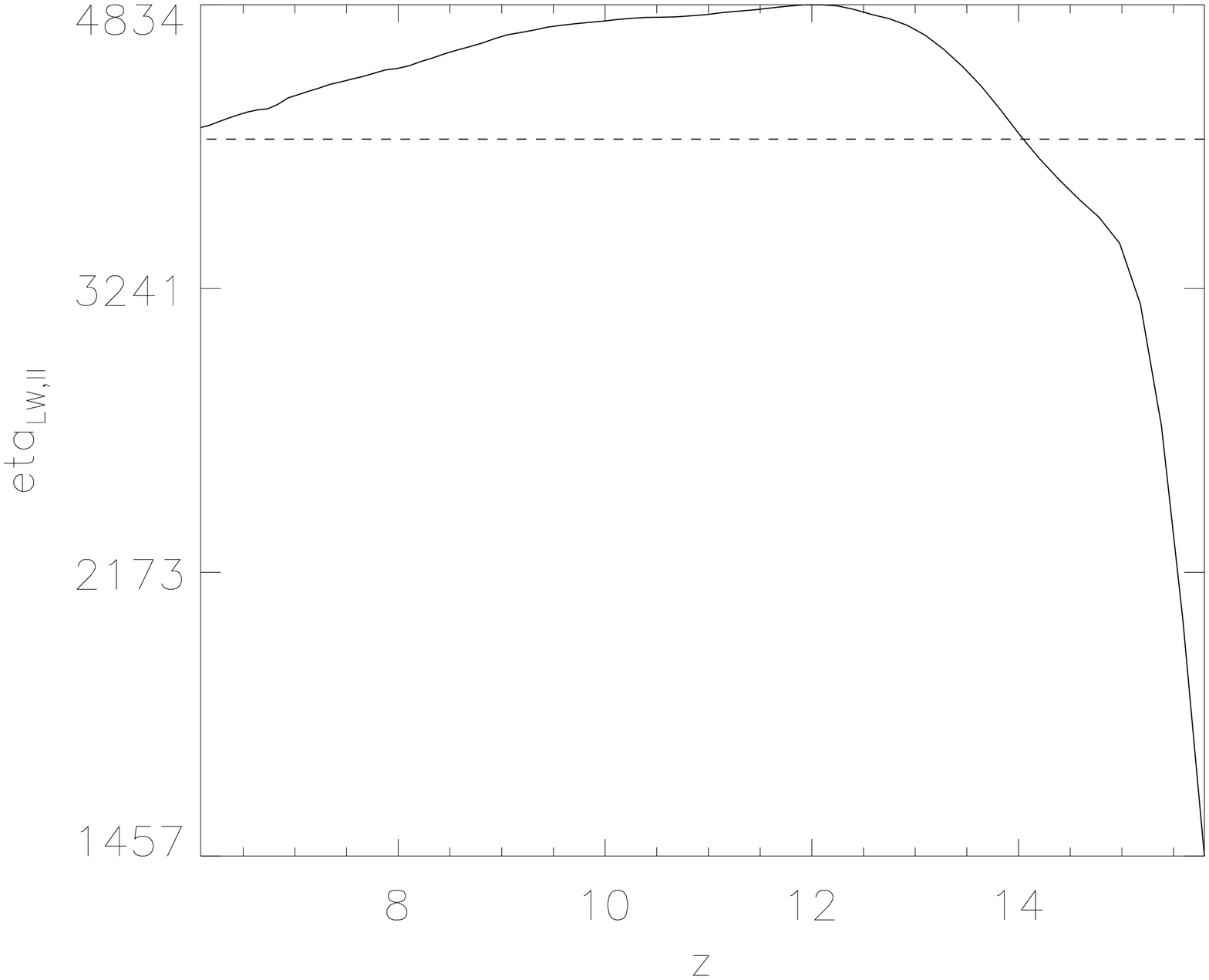}
\caption[$\rm \eta_{LW}$ from Pop II stars]{The cosmic evolution of the $\rm \eta_{LW,II}$ parameter computed on the basis of this work and \cite{Agarwal:2012p2110}. The dashed line corresponds to $\rm \eta_{LW,II}=4\times10^3$ \citep{Greif:2006p99}.}
\label{fig.etaz}
\end{figure}

The spectral parameters are computed self consistently, by identifying all the Pop II stars alive at a given redshift, adding up their SEDs as per their ages (after normalising for the stellar mass, see Eq. \ref{eq.implementkdi} and \ref{eq.implementkde}), and then using Eq. \ref{eq.beta} and \ref{eq.alpha}.
The background level of the LW specific intensity at any given redshift from Pop II stars, J$_{\rm bg,II}$ (in units of 10$^{-21}$ erg/s/cm$^2$/Hz/sr) can be computed using \cite{Greif:2006p99}
\begin{equation}
{\rm J_{bg,II} }= \frac{hc}{4\pi m_H}\eta_{\rm LW,II}\rho_{\rm *,II}(1+z)^3
\end{equation}
where $\eta_{\rm LW,II}$ is the number of LW photons per stellar baryon and $\rho_{\rm *,II}$ is the co--moving mass density of Pop II stars at a given redshift. We compute $\eta_{\rm LW,II}$ self consistently on the basis of the Pop II stellar populations alive at any given redshift (or snapshot $n$ with age $n_t$) using

\begin{equation}
\eta_{\rm LW,II} = \frac{m_H\int\limits_{n_0}^{n_t}{\rm E}_{\rm LW}^{n}dt}{\int\limits_{n_0}^{n_t}\dot M_*^n dt}
\end{equation}
where E$_{\rm LW}^n$ is the total energy output in the LW band (units of erg/s) at each redshift summed over all the stellar populations alive at the given redshift, using their corresponding ages and stellar masses, and $\dot M_*^n$ is the array of star formation rate ($\rm M_{\odot }/yr$). We plot the evolution of $\eta_{\rm LW,II}$ with redshift in Fig. \ref{fig.etaz}, and find that the resultant $\rm J_{bg,II}$ is higher by a factor of $\sim 2$ than the one computed in \cite{Agarwal:2012p2110}.

\begin{figure}
\centering
\includegraphics[width=\columnwidth,]{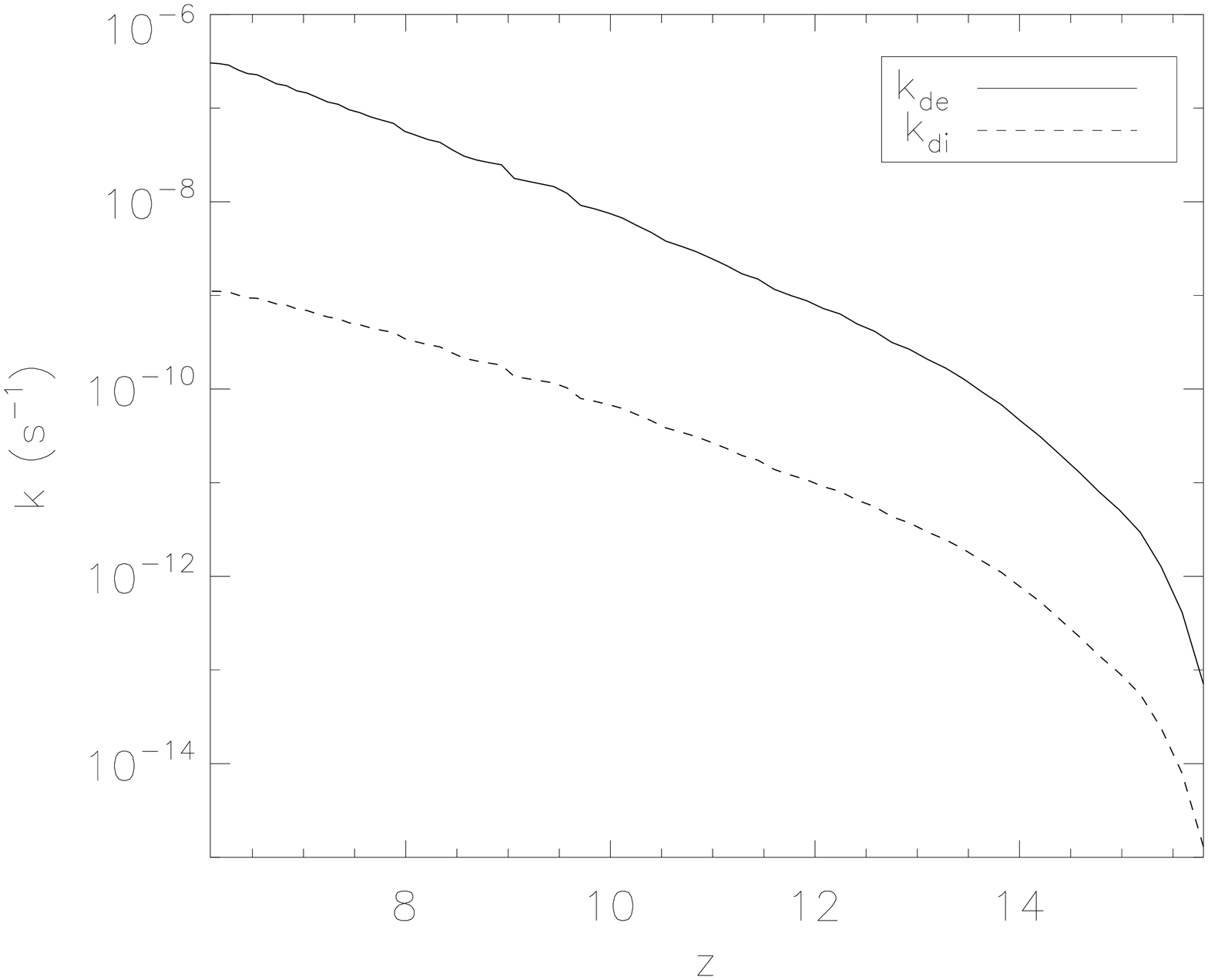}
\caption[Cosmic destruction rates]{Cosmic averaged photo--dissociation and photo--detachment rates computed as a function of redshift using the spectral parameters derived in this study, and J$_{\rm bg, II}$ derived on the basis of \cite{Agarwal:2012p2110}.}
\label{fig.globalk}
\end{figure}

The result of our computation for the global evolution of the reaction rates is plotted in Fig. \ref{fig.globalk}. The reaction rates are highest at later times, due to the role that older stellar populations play in photo--detachment and photo--dissociation. Given that the semi--analytical model of \cite{Agarwal:2012p2110} matches the observational constraints on the CSFR and stellar mass functions, the reaction rate curves represent the average cosmic value of the H$^-$ and H$_2$ destruction in the presence of a LW background emanating from Pop II stars.


\section{Conclusion and Discussion}
\label{sec.conclusions}

In this study, we have demonstrated that using stellar synthesis codes instead of the commonly assumed T5 or T4 thermal black body spectra, to model the SEDs of Pop III and Pop II stellar populations can have a significant impact on thedestruction of H$_2$ and H$^-$. We find that the spectral parameters $\beta$ ($\alpha$) can vary over 5 (12) orders of magnitude depending on the age of the stellar population. The reaction rates themselves however vary only over few orders of magnitude due to the interplay of the spectral parameters with J$_{21}$ that goes into computing the rates. 


It is clear that assuming LW photons are being produced by only young stellar populations with ages 1--10 Myr can lead to an underestimation in the value of the reaction rates. For a given stellar mass, older stellar populations in a continuously star forming galaxy are more efficient in both photo--dissociating H$_2$, and photo--detaching H$^-$, than stars that are assumed to be produced in instantaneous bursts. In the case of photo--detachment of H$^-$, the reaction rate increases with stellar age for the continuous mode, as compared to the bursty mode where it falls off steeply after 10 Myr. For photo--detachment of H$_2$ however, the reaction rate plateaus after 10 Myr for the continuous mode, and again, falls off steeply (more than 10 orders of magnitude) for the bursty mode. Thus it becomes imperative to know the age, star formation history (SFH), and metallicity of the stellar population that is assumed to be the irradiating source, and knowing the value of J$_{21}$ by itself is not enough.

Comparing the burst mode models analysed here to black body curves, we find that photo--dissociation rate of H$_2$ i.e. k$_{\rm di}$, shows little difference at $<10$ Myr for Pop II stars. After 10 Myr however, the T4 photo--dissociation rate drops off rapidly and can be up to 4-5 orders of magnitude lower than the rate derived from stellar synthesis codes. However, the difference is minimal for the case with a continuous SF mode. For the photo--detachment of H$^-$, i.e. k$_{\rm de}$, the difference is quite large and pronounced. At $<50$ Myr, the photo--detachment rates from the T4 spectra are a 3--4 orders of magnitude higher than in the cases with a burst mode of SF, after which the trend reverses and they drop off steeply with respect to the stellar models. For the continuous mode of SF, the T4 photo--detachment rate is 3--4 orders of magnitude higher than that derived from the stellar evolution model at any given age. This demonstrates that one must account for older stellar populations when studying processes that are affected by LW feedback.

When studying the formation of the first stars and galaxies, where LW feedback is critical, reaction rates/rate coefficients derived here could serve as a more accurate prescription than what is available in the literature.
We have derived cosmological average reaction rates, that can serve as self-consistent inputs in high-resolution zoom studies that do not follow the actual star formation history in a large cosmological context.

The newly derived rate coefficients might alter the scenario of direct--collapse black hole formation \citep[e.g.][]{Eisenstein:1995p870,Oh:2002p836,Bromm:2003p22,Koushiappas:2004p871,Lodato:2006p375,Regan:2009p776}, where a critical level of LW specific intensity (J$_{\rm crit}$) is essential in suppressing H$_2$ cooling. Previous studies have done an excellent job at computing the critical level of LW intensity by assuming either a T4 or T5 type spectrum for the irradiating sources \citep{Omukai:2001p128,Shang:2010p33,WolcottGreen:2011p121,Latif:2014p3556}. 
However, repeating their analyses with the reaction rates presented in this study has the potential to significantly increase the value of J$_{\rm crit}$, thus presenting new challenges or avenues to the formation of quasars at $z>6$.

\section*{Acknowledgements}
BA would like to thank Kazu Omukai and Zoltan Haiman for discussions that sparked this project. The authors would like to thank Simon Glover and Andrew Davis for their inputs during the preparation of the manuscript. BA would also like to thank Jan--Pieter Paardekooper and Alessia Longobardi for useful discussions during the early stages of the manuscript. The authors also thank Jonathan Elliott and Laura Morselli for their comments on the draft. 
\bibliographystyle{mn2e}
\bibliography{babib}

\label{lastpage}

\end{document}